\PassOptionsToPackage{table,dvipsnames}{xcolor}

\documentclass[11pt,a4paper,logo]{googledeepmind}

\setleftlogo[120pt]{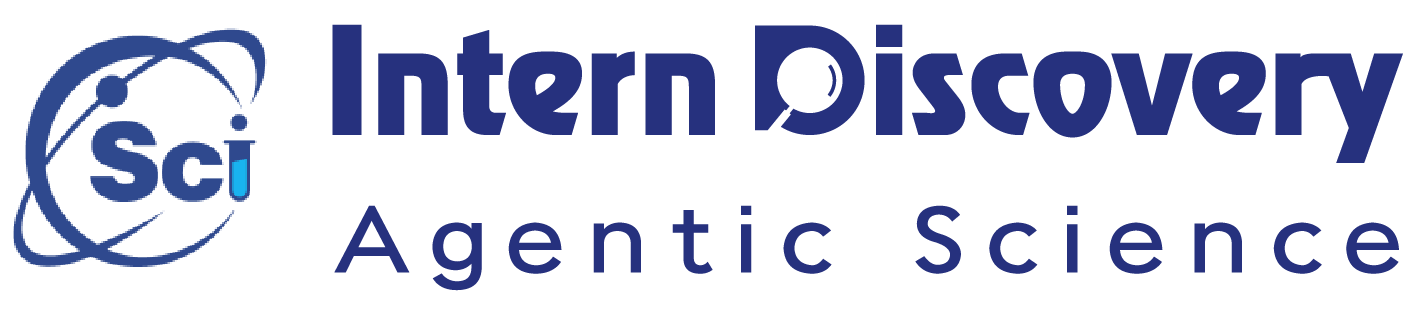}
\setrightlogo[180pt]{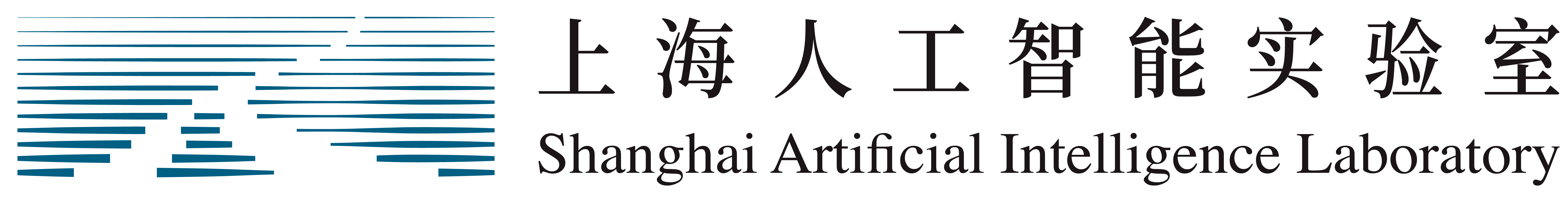}

\usepackage[
    natbib=true,
    backend=biber,
    style=numeric, 
    sorting=none   
]{biblatex}
\AtEveryBibitem{\clearfield{month}}
\AtEveryBibitem{\clearfield{day}}
\usepackage{csquotes}

\newcommand{\BenchName}{PhySciBench}
\newcommand{\AgentName}{DelveAgent}

\usepackage{pdflscape}
\usepackage{pdfpages}
\usepackage{rotating}
\usepackage{setspace}
\usepackage{soul}
\usepackage{subcaption}
\usepackage{cleveref}      
\usepackage[table]{xcolor} 
\sethlcolor{green!14}
\usepackage{array}
\usepackage{multirow}
\usepackage{siunitx}
\usepackage{float}
\usepackage{seqsplit}
\usepackage{framed}
\usepackage{tikz}
\usepackage{listings}
\usepackage{wrapfig}
\usepackage[skins,breakable,most]{tcolorbox}
\usepackage{ragged2e}
\usepackage{makecell}
\usepackage{adjustbox}
\usepackage{etoc}
\usepackage[symbol]{footmisc}
\usepackage{mathrsfs}
\usepackage{placeins}
\usepackage{algorithm}
\usepackage{algorithmicx}
\usepackage{algpseudocode}
\tcbuselibrary{listingsutf8}

\newcolumntype{Y}{>{\RaggedRight\arraybackslash}X}



\definecolor{citecolor}{RGB}{69,123,157}
\hypersetup{
    colorlinks=true,
    linkcolor=citecolor,
    citecolor=citecolor,
    filecolor=black,
    urlcolor=blue
}

\usepackage{tocloft}

\usepackage{etoolbox}
\usepackage{arydshln}
\makeatletter
\patchcmd{\@tocline}
    {\hfil}
    {\leaders\hbox{\hfil}\hfil}
    {}{}
\makeatother

\title{Deep Research in Physical Sciences: A Multi-Agent Framework and Comprehensive Benchmark}

\makeatletter
\renewcommand{\@author}{%
  {\normalfont\bfseries\fontsize{10}{13}\selectfont
   Yigeng Jiang\textsuperscript{1,2,*},
   Tengchao Yang\textsuperscript{1,7,*},
   Taoyong Cui\textsuperscript{1},
   Jiaxing Wan\textsuperscript{3,4},
   Yuan Wang\textsuperscript{6},
   Weida Wang\textsuperscript{1},
   Zhiyu Liu\textsuperscript{3,4},
   Chuyi Peng\textsuperscript{3,4},
   Binzhao Luo\textsuperscript{3},
   Maoli Gao\textsuperscript{3,4},
   Huaihai Huang\textsuperscript{1},
   Yuqianer Zeng\textsuperscript{5},
   Ziyang Zheng\textsuperscript{3,4},
   Dongchen Huang\textsuperscript{3,4},
   Chao Chen\textsuperscript{2},
   Zichao Liu\textsuperscript{1},
   Weiping Shen\textsuperscript{1},
   Shuchen Pu\textsuperscript{1},
   Siyu Zhou\textsuperscript{1},
   Runmin Ma\textsuperscript{1},
   Yusong Hu\textsuperscript{1},
   Fei Chao\textsuperscript{2},
   Bo Zhang\textsuperscript{1},
   Xiawu Zheng\textsuperscript{2},
   Zifu Wang\textsuperscript{1,\textdagger},
   Lei Bai\textsuperscript{1,\textdagger},
   Yunqi Cai\textsuperscript{3,4,\textdagger},
   Shufei Zhang\textsuperscript{1,\textdagger}\par}%
  \vskip0.6em
  {\normalfont\fontsize{8.5}{10.5}\selectfont
   \textsuperscript{1}Shanghai Artificial Intelligence Laboratory\quad
   \textsuperscript{2}Xiamen University\quad
   \textsuperscript{3}Beijing National Laboratory for Condensed Matter Physics and Institute of Physics, Chinese Academy of Sciences\quad
   \textsuperscript{4}Condensed Matter Physics Data Center, Chinese Academy of Sciences\quad
   \textsuperscript{5}University College London\quad
   \textsuperscript{6}Wuhan University\quad
   \textsuperscript{7}Tongji University\par}%
}
\makeatother

\correspondingauthor={\textsuperscript{*}These authors contributed equally. This work was done during their internship at Shanghai Artificial Intelligence Laboratory.\quad\textsuperscript{\textdagger}Corresponding authors.}

\begin{abstract}

Deep research agents are Large Language Model (LLM)-based systems designed for autonomous, multi-step scientific reasoning, and they hold immense potential for accelerating research in the physical sciences. However, comprehensive and in-depth evaluations of their capabilities within this domain remain lacking. To address this gap, we introduce \textbf{\BenchName}, a benchmark highly relevant to physical science research, comprising 200 expert-curated questions, balanced between physics and chemistry, across six task categories that reflect real-world scientific workflows. Evaluations of state-of-the-art models and agent systems on \textbf{\BenchName} reveal limited performance; even the strongest baseline, Gemini Deep Research, achieves an accuracy of only 33.5\%. Analysis of failure cases identifies three recurrent deficiencies: fragility in extended reasoning chains, limited knowledge transfer across steps, and a lack of physics-grounded self-verification. Motivated by these findings, we develop \textbf{\AgentName}, a modular multi-agent framework equipped with an adaptive planning loop, dual-granularity memory, and a hierarchical physics-grounded reflection mechanism. Across four scientific benchmarks, \textbf{\AgentName} improves accuracy by up to 7.5 percentage points while reducing inference costs to approximately one-third of the strongest baseline. These results establish the significance of \textbf{\BenchName} as a critical benchmark for evaluating AI systems in the physical sciences and demonstrate that architectural specialization can effectively enhance the reliability of autonomous scientific research. Our data and code are publicly available at \url{https://github.com/yigengjiang/physci-deepresearch}.
\end{abstract}

\begin{document}
\sloppy
\maketitle

\section{Introduction}\label{sec1}


The research paradigm in the physical sciences is highly dependent on the profound integration of diverse scientific evidence. These sources span a broad spectrum, encompassing experimental measurements, scientific figures, structured materials data, and the primary literature~\cite{jumper2021alphafold, merchant2023scaling, szymanski2023alab, jain2013mp, zhao2024spacia}. Concurrently, this integration process necessitates a deep understanding of vast knowledge repositories, complex synthesis and reasoning, and the efficient utilization of external tools. However, driven by the explosive growth of high-throughput experimental data and the ever-expanding volume of scientific literature, the traditional model of knowledge integration, which relies heavily on individual researchers, has reached a critical bottleneck. This limitation underscores an urgent need for artificial intelligence systems capable of autonomously executing multi-step scientific reasoning~\cite{boiko2023coscientist, bran2024chemcrow}.

Deep research agents have emerged as a highly promising framework to overcome the limitation for multi-step scientific reasoning. Built upon Large Language Models (LLMs), these systems can autonomously decompose complex problems, flexibly invoke external tools, and dynamically synthesize diverse evidence across multi-step interactions~\cite{yao2023react, wei2022cot, openai2025deepresearch, google2025geminideepresearch}. Recent studies indicate that such agentic workflows have already demonstrated preliminary capabilities in supporting core tasks within the physical sciences, including multi-step problem-solving, scientific hypothesis generation, structure–property correlation analysis, and experimental design~\cite{feng2026autonomous, woodruff2026accelerating, gottweis2025coscientist, lu2024aiscientist, boiko2023coscientist, bran2024chemcrow, miao2025physmaster, qiu2025physics, hellert2025agentic, wei2025aiscience, pieropan2025neurons}. These advancements provide compelling evidence that agentic LLM systems are transcending the traditional static question-answering paradigm, steadily advancing toward a new era of open-ended and profound autonomous scientific reasoning.

Nevertheless, existing benchmarks remain inadequate for evaluating the capability of agents to undertake authentic research tasks in the physical sciences. On the one hand, domain-specific benchmarks in chemistry and the physical sciences~\cite{mirza2025chembench, laurent2024labbench, tian2024scicode, chen2024scienceagentbench, qiu2025phybench, wang2026cmphysbench, pan2026cmt} are not only scarce but also rarely address core real-world research scenarios. These tests typically evaluate single tasks in isolation, lacking the assessment of extended workflows integrated with external tools. On the other hand, general-purpose scientific evaluation standards, such as the closed-ended HLE~\cite{phan2025hle} and the deep-research-oriented yet text-only SGI-Bench~\cite{sgibench} and FrontierScience~\cite{frontiersci}, fail to encompass the multimodal parsing, structured data extraction, and code execution essential for genuine scientific inquiry. These multimodal, coding and tool-use capabilities, however, constitute the foundational pillars of practical research in the physical sciences.

To overcome the above limitations, we introduce \textbf{\BenchName}, a benchmark specifically tailored to evaluate research capabilities within the physical sciences. Distinct from prior evaluations, \textbf{\BenchName} comprehensively covers three cognitive stages: information extraction, scientific understanding, and application and creation, thereby fully capturing the end-to-end workflow through which deep research agents empower the physical sciences. This benchmark comprises 200 expertly curated questions spanning six task categories: multimodal question answering, long-context question answering, structured information extraction, scientific reasoning, experimental design, and code generation, which faithfully replicate common physical science workflows. Drawn extensively from authentic figures in the primary literature and professionally adapted textbook exercises, these questions are designed to rigorously assess model performance in multi-step tool utilization and cross-source reasoning. Furthermore, all questions are authored directly by domain experts to stringently maintain scientific authenticity, ensuring a profound alignment between the evaluation tasks and actual research demands.

When we evaluate state-of-the-art general-purpose agents on \textbf{\BenchName}, performance remains limited: the strongest baseline, Gemini Deep Research~\cite{google2025geminideepresearch}, achieves 33.5\% accuracy.
Analysis of 326 failure cases reveals three recurring limitations: brittle long-horizon reasoning, weak transfer of domain experience across steps and insufficient physics-grounded verification. These failures arise when agents do not revise plans after intermediate errors, do not reuse task-relevant strategies across problems, and do not validate intermediate outputs against physical constraints. Together, these observations suggest that improving performance in physical sciences will require architectural specialization rather than scale alone. 

Motivated by these observations, we accordingly propose \textbf{\AgentName}, a modular multi-agent framework that combines adaptive planning loop, dual-granularity memory, and hierarchical physics-grounded reflection.
In this framework, adaptive planning enables replanning after receiving intermediate outcomes, memory supports reuse of task experience and domain knowledge, and physics-grounded reflection checks intermediate and final outputs for scientific consistency. 
Across \textbf{\BenchName} and three additional scientific reasoning benchmarks, \textbf{\AgentName} improves accuracy by up to 7.5 percentage points over the strongest baseline while reducing inference cost to about one-third of Gemini Deep Research; ablation studies support the contribution of each component.
Together, these results position \textbf{\BenchName} as a benchmark for evaluating AI systems in the physical sciences and identify \textbf{\AgentName} as a practical step towards more reliable autonomous scientific research.
\section{Results}\label{sec2}

\begin{figure*}[!htbp]
    \centering
    \includegraphics[width=\linewidth]{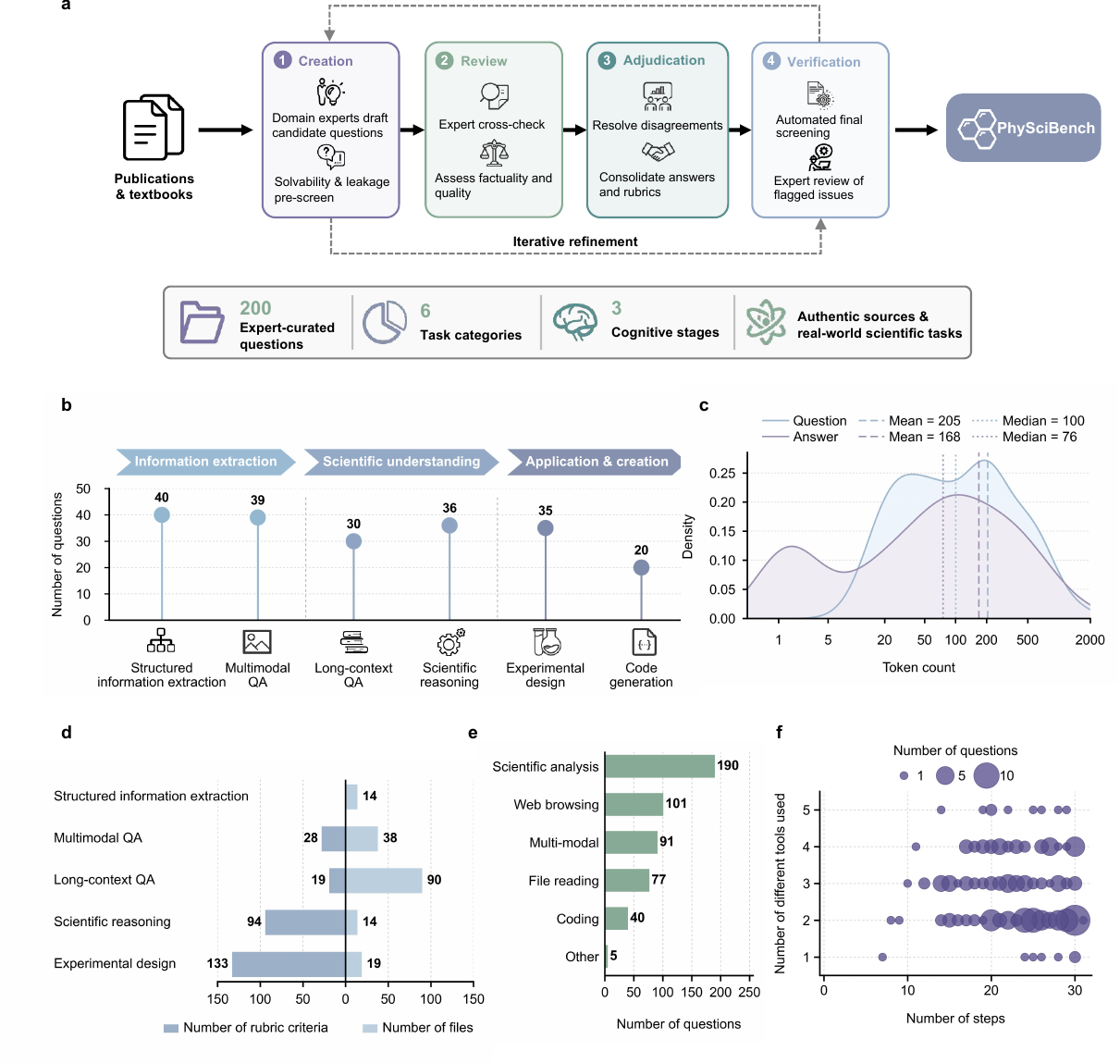}
    \caption{\textbf{Construction and statistical characterisation of {\BenchName}.}
    \textbf{a},~Multi-stage construction pipeline: expert creation, peer review, adjudication and automated leakage screening, with iterative refinement returning failing items to earlier stages.
    \textbf{b},~Per-category question counts grouped by cognitive stage (Information extraction, Scientific understanding, Application \& creation).
    \textbf{c},~Distributions of question and answer token counts on a log-token axis.
    \textbf{d},~Per-category number of supporting input files (light blue) and rubric criteria (dark blue).
    \textbf{e},~Number of questions requiring each of six tool-capability classes (a question may require multiple classes).
    \textbf{f},~Joint distribution of solution-step count and number of distinct tool classes per question; marker area is proportional to question count at each coordinate.}
    \label{fig:dataset_overview}
\end{figure*}

\subsection{Construction of {\BenchName}}\label{subsec:dataset_introduction}

\begin{figure*}[!htbp]
    \centering
    \includegraphics[width=\linewidth]{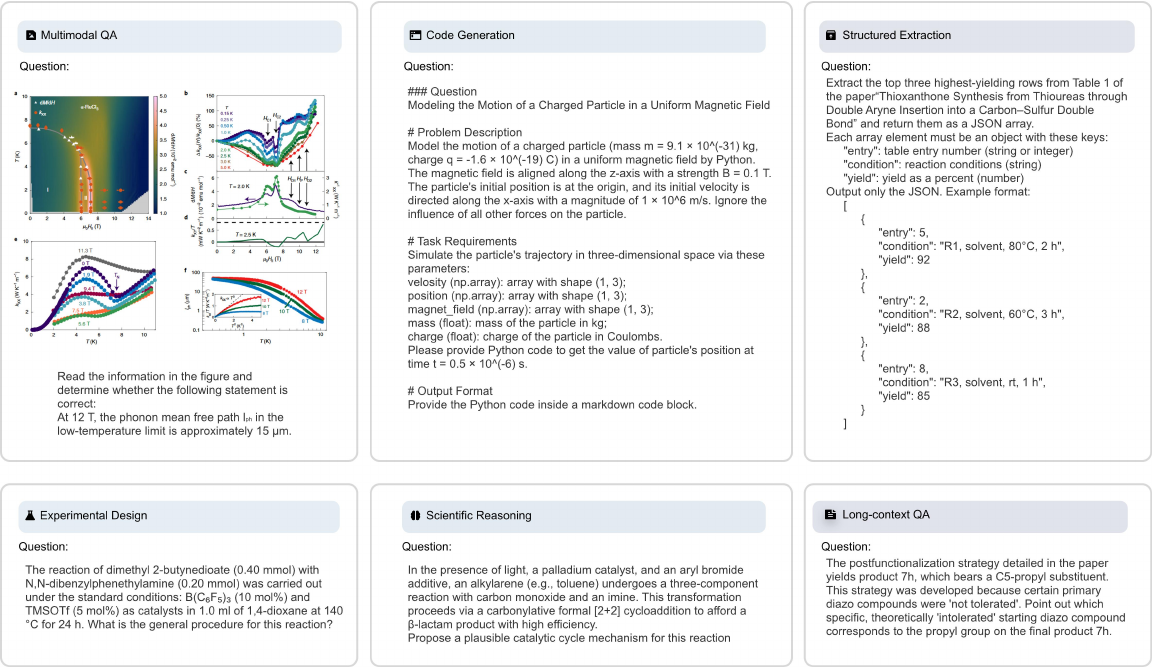}
    \caption{\textbf{Representative {\BenchName} questions from all six task categories.} One expert-curated item per category (Multimodal QA, Code Generation, Structured Information Extraction, Experimental Design, Scientific Reasoning, Long-context QA), illustrating the range of source materials (experimental figures, tabulated data, research articles, methods sections) and answer formats (atomic values, structured records, executable code, free-form scientific prose).}
    \label{fig:example}
\end{figure*}

To bridge the gap between standardized evaluation and authentic scientific inquiry, we introduce {\BenchName}, a comprehensive dataset comprising 200 high-quality questions derived from recent high-impact physical-science publications and adapted from textbook problems; all questions are originally crafted by domain experts to mitigate training-data contamination. The construction pipeline integrates expert creation, peer cross-review, adjudication of disagreements and an automated leakage-screening pass with expert verification of flagged items, with iterative refinement returning failing questions to earlier stages (Figure~\ref{fig:dataset_overview}a). Crucially, to faithfully mirror the genuine cognitive workflow of physical-science research, we systematically stratify the six task categories into three progressive cognitive stages, with two tasks per stage. \textbf{Information extraction} (comprising Multimodal QA and Structured Information Extraction) probes whether the agent can perceive scientific figures and parse unstructured material into discrete answers or schema-conformant records. \textbf{Scientific understanding} (comprising Long-context QA and Scientific Reasoning) probes whether the agent can integrate evidence across documents and derive principle-grounded conclusions. \textbf{Application and creation} (comprising Experimental Design and Code Generation) probes whether the agent can synthesise new scientific artefacts, from procedurally complete protocols to executable computational models. The six categories are near-equally balanced (Figure~\ref{fig:dataset_overview}b). The benchmark is balanced by design across physics and chemistry (100 questions each), and each question is mapped to a subfield drawn from established community classifications (PhySH and Physical Review sections for physics; ACS and IUPAC divisions for chemistry); the resulting coverage centres on condensed-matter physics and on synthetic organic chemistry and catalysis. Question and answer token counts are bimodally distributed with question text longer on average (mean 205, median 100) than answer text (mean 168, median 76; Figure~\ref{fig:dataset_overview}c). Long-context QA carries the largest input-file load and Experimental Design the densest rubric (Figure~\ref{fig:dataset_overview}d), and questions span six broad tool-capability classes dominated by scientific analysis (Figure~\ref{fig:dataset_overview}e). The bulk of the benchmark concentrates between 20--30 reasoning steps and 2--4 tool classes, indicating that {\BenchName} primarily evaluates extended, multi-tool scientific workflows (Figure~\ref{fig:dataset_overview}f). A representative item from each task category is shown in Figure~\ref{fig:example}.

To facilitate reliable assessment across these diverse answer formats, we developed an automated evaluation framework that integrates exact-match normalisation, rule-based syntax and key/value checks, rubric-guided LLM judgement and sandboxed code execution (Figure~\ref{fig:model_analysis}a). In a blinded validation against manual grading by two domain experts (one for physics and one for chemistry, each grading items within their respective discipline), the composite pipeline showed strong agreement with expert scores (Spearman $\rho = 0.80$, 95\% CI [0.75, 0.85], $p < 10^{-10}$, $n = 196$ paired items; Figure~\ref{fig:model_analysis}b), substantially exceeding a single-LLM-judge baseline ($\rho = 0.57$, 95\% CI [0.47, 0.66]), lexical metrics (ROUGE-2 $\rho = 0.31$; BLEU $\rho = 0.21$) and embedding-based BERTScore-F1 ($\rho = 0.15$, 95\% CI [0.01, 0.28]). These results indicate that the evaluation pipeline provides a practical proxy for expert assessment.

\subsection{Substantial headroom remains for generalist agents on physical-science deep research}\label{subsec:failure_analysis}

\begin{figure*}[!htbp]
    \centering
    \includegraphics[width=\linewidth]{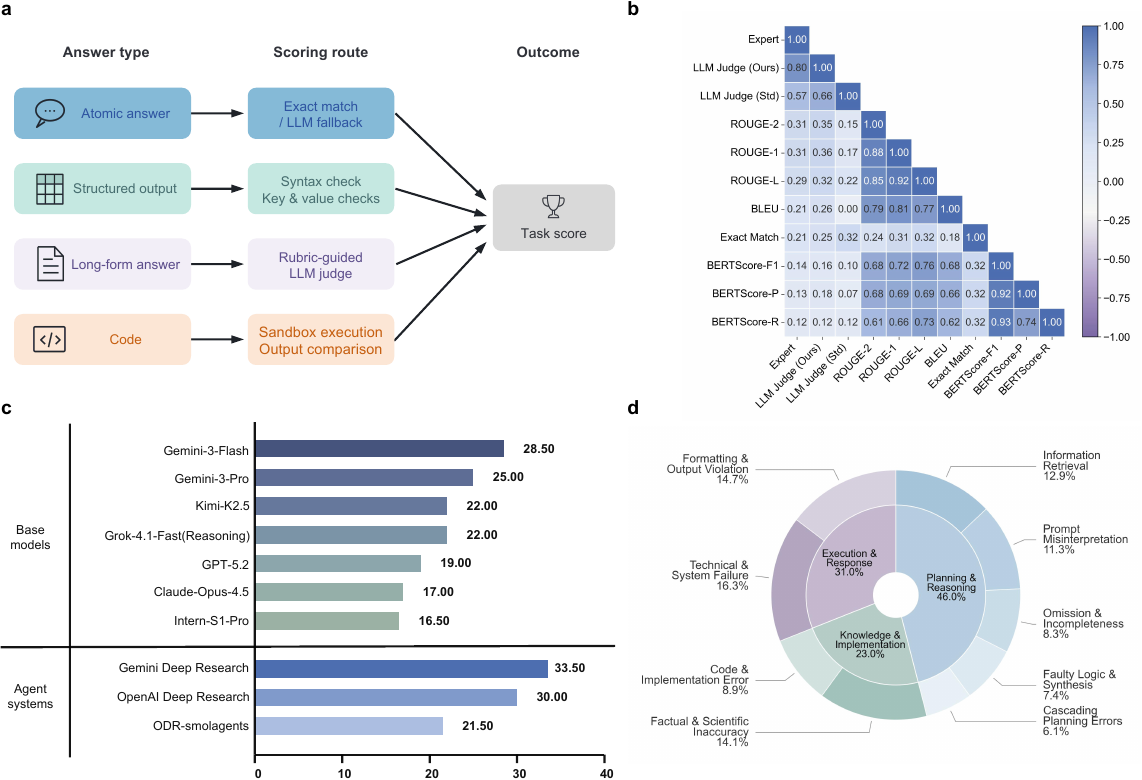}
    \caption{\textbf{Evaluation protocol and failure profile of generalist agents on {\BenchName}.}
    \textbf{a},~Composite scoring pipeline combining exact-match grading with LLM-judge fallback, rule-based syntactic and key/value checks, rubric-guided LLM judgement and sandboxed code execution.
    \textbf{b},~Lower-triangular Spearman rank-correlation matrix among an independent expert rubric score (continuous 0--1; obtained from blinded grading by two domain experts---one for physics and one for chemistry, each grading items within their respective discipline) and ten evaluation metrics: a single-LLM-judge baseline, the composite pipeline (this work), exact-match, ROUGE-1/2/L, BLEU, and BERTScore (precision/recall/F1). Cells annotate Spearman $\rho$. $n = 196$ paired items; the four structured-information-extraction items for which lexical metrics are undefined are excluded.
    \textbf{c},~Overall accuracy on {\BenchName} of seven base models and three generalist agent systems.
    \textbf{d},~Failure-mode taxonomy from 326 verified errors of the strongest baseline, grouped into three macro-categories (planning and reasoning, execution and response, knowledge and implementation) with fine-grained sub-types on the outer ring.}
    \label{fig:model_analysis}
\end{figure*}


Existing generalist large language models exhibit a substantial performance gap on {\BenchName}. The top-performing baseline, Gemini Deep Research~\cite{google2025geminideepresearch}, achieved an accuracy of only 33.5\%; all evaluated base models scored below 30\%, and reasoning-enhanced systems such as Gemini~3~Flash~\cite{google2025gemini3flash} and Grok-4.1-Fast~\cite{xai2026grok41fast} scored lower still (Figure~\ref{fig:model_analysis}c), indicating that physical-science tasks require capabilities beyond standard chain-of-thought processing.

To characterise these failures, we analysed 326 verified error cases from the strongest baseline, Gemini Deep Research (Figure~\ref{fig:model_analysis}d). We identified three recurrent failure families: planning and reasoning failures, knowledge and implementation failures, and execution and response failures. We use these categories as diagnostic groupings to relate benchmark failures to the design of {\AgentName}.

\begin{itemize}
\item \textbf{Planning and reasoning failures accounted for 46\% of cases.}
The largest fraction of question-level failures arises from unstable long-horizon reasoning trajectories. In these cases, early errors in task interpretation, evidence retrieval, or intermediate synthesis frequently propagate to downstream steps, resulting in incomplete or incorrect final solutions. This pattern motivates our planning module, which is designed to support dynamic replanning by reassessing intermediate outputs and revising subsequent actions when the current trajectory no longer aligns with the task requirements.

\item \textbf{Knowledge and implementation failures accounted for 23\% of errors.}
A substantial subset of failures reflects limitations in reusing domain knowledge and prior problem-solving experience. Typical cases include scientifically inaccurate statements, hallucinated mechanisms or parameters, and code or formula implementations that do not faithfully realise the intended physical or mathematical model. This pattern motivates our memory module, which accumulates and retrieves structured scientific knowledge and successful prior trajectories to support more grounded and reusable solution construction.

\item \textbf{Execution and response failures accounted for 31\% of errors.}
Another major fraction of failures originates from breakdowns in task execution and answer delivery rather than from scientific reasoning alone. These failures include toolchain timeouts, document or multimodal processing failures, and violations of required output formats. This pattern motivates our reflection module, which monitors intermediate execution states, verifies whether outputs satisfy task constraints, and triggers corrective actions such as retrying, reformatting, or revising incomplete responses.
\end{itemize}

Together, these failure modes suggest that the performance ceiling of current deep-research systems in physical science is architectural rather than purely a matter of scale. The dominant deficits are weak replanning, poor transfer of prior experience, and insufficient physics-grounded verification. We therefore designed \textbf{\AgentName} around three corresponding components: an \textit{Adaptive Planning Loop}, \textit{Dual-Granularity Memory} and \textit{Hierarchical Physics-Grounded Reflection}.

\subsection{{\AgentName} addresses systemic deficiencies}\label{subsec:benchmark_results}

\begin{figure*}[!htbp]
    \centering
    \includegraphics[width=\linewidth]{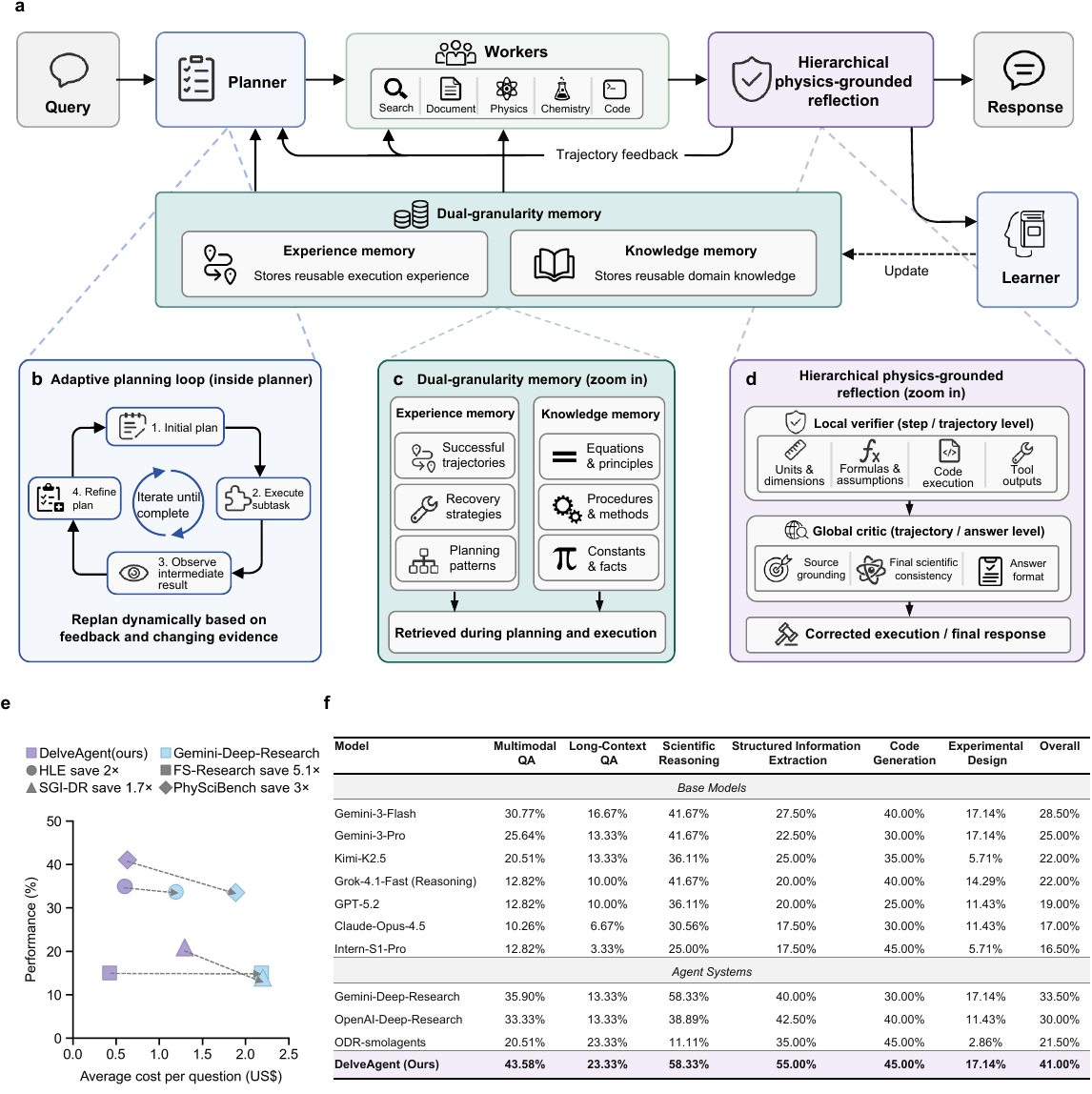}
    \caption{\textbf{{\AgentName} architecture and overall performance.}
    \textbf{a},~End-to-end workflow: a central Planner dispatches subtasks to specialised Workers; Worker outputs pass through a hierarchical physics-grounded reflection block before response assembly. A dual-granularity memory is shared between Planner and Workers, and a Learner module distils verified trajectories from the reflection block back into the knowledge memory.
    \textbf{b},~Adaptive planning loop: initial plan $\to$ subtask execution $\to$ intermediate observation $\to$ plan refinement.
    \textbf{c},~Dual-granularity memory combining experience memory (successful trajectories, recovery strategies, planning patterns) and knowledge memory (equations, procedures, physical constants).
    \textbf{d},~Hierarchical physics-grounded reflection: a local verifier (units, dimensions, formulas, code, tools) and a global critic (source grounding, scientific consistency, format).
    \textbf{e},~Cost--accuracy comparison on four scientific benchmarks, with relative cost savings annotated.
    \textbf{f},~Per-category accuracy on {\BenchName} for seven base models, three generalist agent systems and {\AgentName}.}
    \label{fig:results_overview}
\end{figure*}

The {\AgentName} framework (Figure~\ref{fig:results_overview}a), whose three architectural components directly target the deficiencies identified above, was evaluated on {\BenchName} alongside all baselines. The Planner orchestrates a panel of specialised Workers and, together with the Workers, queries a dual-granularity memory shared between them; Worker outputs are routed through a hierarchical physics-grounded reflection block before assembly into the final response. The adaptive planning loop iterates initial-plan to subtask execution to intermediate-result observation to plan refinement, replanning dynamically rather than committing to an initial decomposition (Figure~\ref{fig:results_overview}b). The dual-granularity memory pairs an experience store of successful trajectories, recovery strategies and planning patterns with a knowledge store of equations, procedures and physical constants, both accessed during planning and worker execution (Figure~\ref{fig:results_overview}c). Hierarchical physics-grounded reflection couples a step-level local verifier (units, dimensions, formulas, code execution and tool outputs)~\cite{liang2025saiunit} with a trajectory-level global critic (source grounding, scientific consistency, format compliance), with corrective execution triggered before the final response (Figure~\ref{fig:results_overview}d).

As illustrated in the per-category breakdown (Figure~\ref{fig:results_overview}f), {\AgentName} achieved an overall accuracy of 41.0\% on {\BenchName}, a +7.5 percentage-point improvement over the strongest baseline, Gemini Deep Research (33.5\%). This absolute accuracy reflects the deliberate design of {\BenchName} as an unsaturated evaluation instrument: expert-originated questions, anti-contamination safeguards and rigorous difficulty calibration collectively preserve substantial headroom, ensuring that the benchmark discriminates among systems rather than saturating at current capability levels. The largest gains emerged in Structured Information Extraction (+15.0~p.p.), where the Dual-Granularity Memory supplies format templates that reduce schema errors, and in Code Generation (+15.0~p.p.), where the Adaptive Planning Loop enables iterative debugging. Multimodal QA improved by +7.7~p.p., reflecting the benefit of physics-grounded verification on visual interpretation tasks. Performance on Scientific Reasoning matched the strongest baseline (58.3\%), and Long-context QA improved by +10.0~p.p.\ over Gemini Deep Research, reflecting the benefit of the dual-granularity memory for cross-document evidence integration. Experimental Design plateaued at the strongest-baseline level (17.1\%), the one category in which {\AgentName} did not exceed Gemini Deep Research. Notably, {\AgentName} achieves these improvements at a small fraction of the inference cost of Gemini Deep Research across all four scientific benchmarks (Figure~\ref{fig:results_overview}e).

Ablation studies quantify the contribution of individual components. Disabling the Dual-Granularity Memory or the Hierarchical Physics-Grounded Reflection each reduced accuracy by 3.0~p.p., and disabling the Adaptive Planning Loop reduced accuracy by 2.5~p.p. The three single-component drops are of comparable magnitude on the 200-item benchmark, and we describe them as comparable contributions rather than identifying any one component as dominant. Per-category attributions follow the design intent of each module: removal of the memory module coincided with the largest drops in Scientific Reasoning and Experimental Design (tasks that rely most heavily on accumulated domain knowledge and transferable planning strategies), whereas removal of reflection coincided with the largest drops in execution-intensive categories, where intermediate verification can intercept errors before they propagate. Removing all three components simultaneously produced a 6.5~p.p.\ degradation, less than the sum of the individual ablations (8.5~p.p.) but substantially larger than any single removal, consistent with partial functional overlap whereby the three modules can mutually compensate. The stripped-down configuration narrows the gap with the strongest generalist baseline, Gemini Deep Research, from 7.5~p.p.\ to 1.0~p.p., supporting the interpretation that the three architectural components, rather than the underlying foundation model, account for the bulk of the {\AgentName} improvement.

Qualitative comparison of failure profiles between {\AgentName} and the baselines reveals a fundamental shift in the dominant failure mode. In baseline systems, errors concentrate in coarse-grained agentic instabilities (cascading planning failures, tool-execution timeouts and wholesale format violations) that reflect upstream architectural fragility. The joint action of planning, memory and reflection substantially eliminates these architectural failure patterns, advancing the residual error distribution toward finer-grained scientific grounding and fidelity challenges. Analysis of 100 sampled {\AgentName} failure cases identifies four principal downstream bottlenecks: evidence binding errors in which data are attributed to incorrect experimental conditions, constraint satisfaction failures where outputs violate explicit task specifications, formal-specification fidelity issues including schema mismatches and numerical precision loss, and source disambiguation difficulties when multiple documents present conflicting information. A subset of these bottlenecks (particularly constraint satisfaction and formal-specification fidelity) remains amenable to reduction through stronger typed orchestration and validation gates. The remainder, concentrated in specialised scientific figure interpretation and cross-document synthesis, reflects genuine capability boundaries of current multimodal foundation models~\cite{alampara2025probing,liutowards}.

\subsection{Generalization and Robustness}\label{subsec:generalization}

\begin{table}[!htbp]
\centering
\caption{\textbf{Generalization performance on public benchmarks.} Models lacking the required multi-modal capabilities (e.g., PDF or image processing) for specific benchmarks are denoted as N/A.}\label{tab:public_benchmark_generalization}
\begin{tabular}{lcccc}
\toprule
\textbf{Model} &
\makecell[c]{\textbf{HLE} \\ \textbf{(Phy/Chem)}} &
\makecell[c]{\textbf{SGI-DR} \\ \textbf{(Phy/Chem)}} &
\makecell[c]{\textbf{FS-Research} \\ \textbf{(Phy/Chem)}} &
\textbf{Average} \\
\midrule
\multicolumn{5}{c}{\textit{Base Models}} \\
\midrule
Qwen3-VL-Plus                  & 8.35\%   & 6.98\%   & 0.00\%   & 5.11\%  \\
GPT-5.2                        & 9.37\%   & 11.63\%   & 10.00\%  & 10.33\%  \\
Grok-4.1-Fast (Reasoning)      & 11.39\%  & 6.98\%   & 12.50\%  & 10.29\% \\
Claude-Opus-4.5                & 11.90\%  & 9.30\%   & 7.50\%   & 9.57\%  \\
Kimi-K2.5                      & 19.24\%  & 6.98\%   & 12.50\%  & 12.91\% \\
DeepSeek-v3.2-Speciale         & N/A      & 4.65\%   & 10.00\%  & 7.33\%$^\dagger$  \\
Intern-S1-Pro                  & 7.34\%   & 6.98\%   & 10.00\%  & 8.11\% \\
Gemini-3-Pro                   & 33.16\%  & 6.98\%   & 12.50\%  & 17.55\% \\
Gemini-3-Flash                 & 32.91\%  & 11.63\%  & 5.00\%  & 16.51\% \\
\midrule
\multicolumn{5}{c}{\textit{Agent Systems}} \\
\midrule
OpenAI Deep Research           & 12.66\%  & 2.33\%   & 10.00\%  & 8.33\%  \\
Gemini Deep Research           & 33.67\%  & 13.95\%  & 15.00\%  & 20.87\% \\
ODR-smolagents                 & 5.57\%   & 11.63\%  & 0.00\%   & 5.73\%  \\
X-Master                       & N/A      & 2.33\%   & 2.50\%   & 2.42\%$^\dagger$  \\
Tongyi DeepResearch            & N/A      & 2.33\%   & 5.00\%  & 3.67\%$^\dagger$  \\
\midrule
\textbf{\AgentName{} (Ours)}   & \textbf{34.94\%} & \textbf{20.93\%} & \textbf{15.00\%} & \textbf{23.62\%} \\
\bottomrule
\end{tabular}

{\footnotesize $^\dagger$Average computed over available (non-N/A) benchmarks only.}
\end{table}

To assess generalisation beyond {\BenchName}, we evaluated {\AgentName} on three established public benchmarks: HLE~\cite{phan2025hle}, a closed-ended expert-level evaluation comprising 395 physics and chemistry questions designed to resist retrieval-based strategies; SGI-DR~\cite{sgibench}, a deep-research track of 43 expert-curated questions requiring iterative multi-step reasoning over complex scientific content; and FS-Research~\cite{frontiersci}, an open-ended research track of 40 tasks graded by a 10-point rubric emphasising scientific judgement.

As presented in Table~\ref{tab:public_benchmark_generalization}, {\AgentName} achieved the highest average accuracy across the three benchmarks (23.62\%), outperforming the strongest baseline, Gemini Deep Research (20.87\%). The largest improvement emerged on SGI-DR (+6.98 percentage points), where the Adaptive Planning Loop and Dual-Granularity Memory directly support the iterative evidence-gathering workflow that this benchmark demands. On HLE, {\AgentName} achieved 34.94\% versus 33.67\% for Gemini Deep Research, a modest gain consistent with HLE's emphasis on closed-ended factual recall where architectural innovations provide limited leverage. Performance on FS-Research was comparable to the strongest baseline, indicating that the gains extend to the most open-ended setting.

\subsection{Performance on real-world scientific tasks}\label{subsec:real_world_application}

\begin{figure*}[!htbp]
    \centering
    \includegraphics[width=\linewidth]{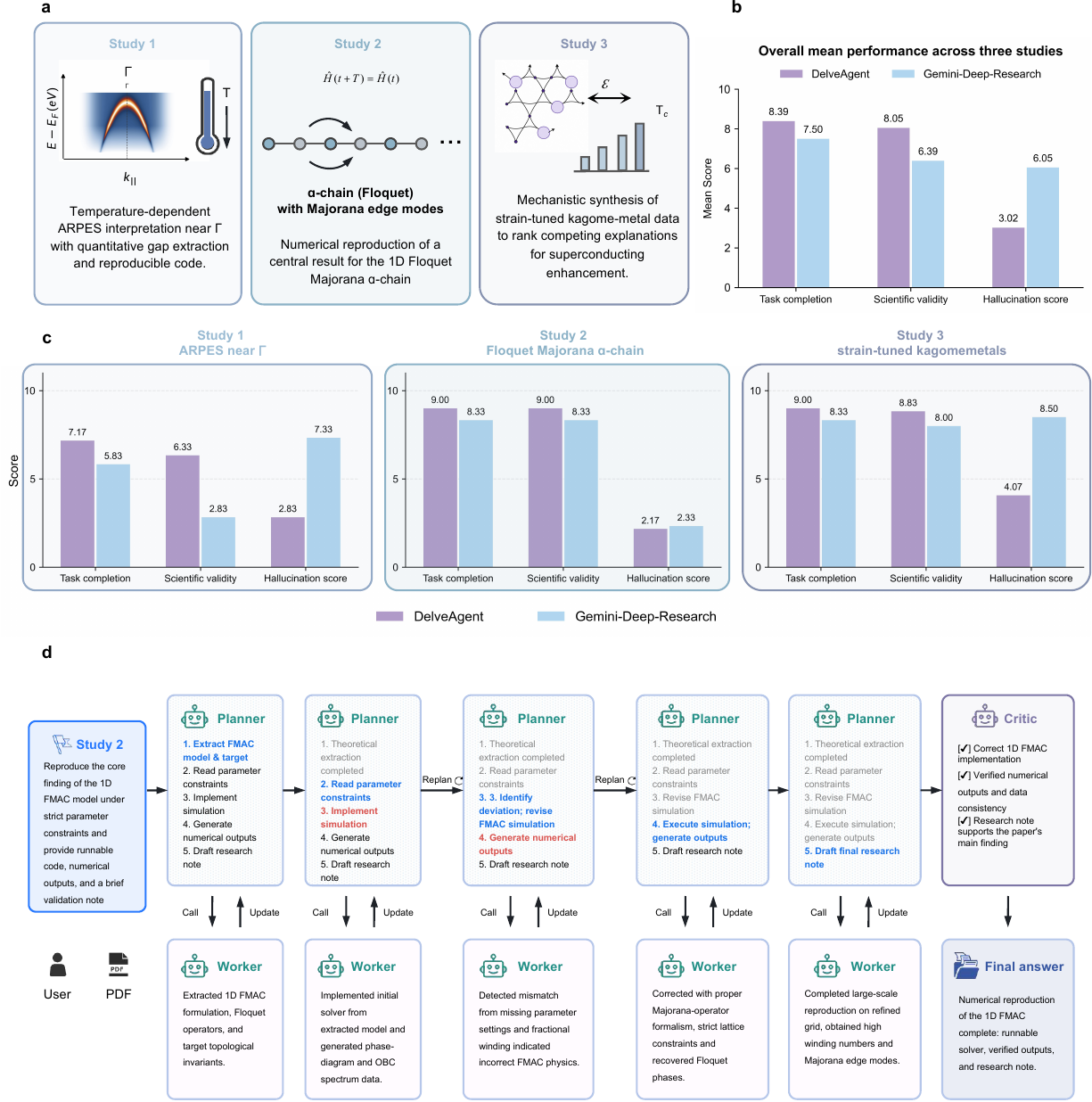}
    \caption{\textbf{End-to-end performance on three expert-designed real-world physical-science tasks.}
    \textbf{a},~Task setups. Study~1: temperature-dependent ARPES with quantitative gap extraction; Study~2: numerical reproduction of the 1D Floquet Majorana $\alpha$-chain; Study~3: mechanistic synthesis of strain-tuned kagome-metal data.
    \textbf{b},~Mean scores across the three studies on three rubric-based 0--10 dimensions (task completion, scientific validity, hallucination score; lower hallucination is better), graded blind by three independent domain experts per task.
    \textbf{c},~Per-study breakdown of the three dimensions.
    \textbf{d},~Representative {\AgentName} execution trajectory on Study~2, showing Planner--Worker alternation, mid-trajectory replanning after a worker detected fractional winding indices signalling an incorrect FMAC implementation, and final Critic verification.}
    \label{fig:real_world}
\end{figure*}

To assess end-to-end performance on realistic physical-science workflows, three physical-science domain experts each designed an open-ended, single-turn research task drawn from their own active research area: rapid interpretation of temperature-dependent ARPES data near $\bar{\Gamma}$ with quantitative gap extraction and reproducible code (Study~1), numerical reproduction of a central result for the 1D Floquet Majorana $\alpha$-chain (FMAC) under prescribed parameter constraints (Study~2), and mechanistic synthesis of strain-tuned kagome-metal data ranking three competing explanations for superconducting enhancement (Study~3); the three task setups are summarised in Figure~\ref{fig:real_world}a. Each task was supplied to {\AgentName} and to Gemini Deep Research as a closed packet of figures, CSV files and methods notes; the verbatim model outputs were then graded by three independent physical-science experts under a blind protocol, with final scores reported as the per-task mean across reviewers.

Each response was scored on three rubric-based 0--10 dimensions. \textbf{Task completion} measures whether the response produces every artefact and answer field requested in the task statement (numerical estimates with uncertainty, structured sections, runnable code, output files). \textbf{Scientific validity} measures whether the conclusion is grounded in the supplied evidence with appropriate operational definitions, cross-validation between image- and CSV-level observations, and explicit uncertainty treatment. \textbf{Hallucination score} (lower is better) penalises external-source leakage, fabricated quantitative values, overclaims about mechanism or symmetry, and presentation of unsupported speculation as established fact.

Across the three studies, {\AgentName} attained higher means than Gemini Deep Research on every dimension: task completion 8.39 vs 7.50, scientific validity 8.05 vs 6.39, and hallucination score 3.02 vs 6.05 (lower is better; Figure~\ref{fig:real_world}b). The aggregate gap is dominated by the two open-ended studies that demand a tight coupling between supplied evidence and the final claim (Figure~\ref{fig:real_world}c). In Study~1 (ARPES), Gemini Deep Research returned an inconclusive verdict and recommended additional measurements rather than committing to a Dirac-cone interpretation, receiving a scientific-validity score of 2.83 against {\AgentName}'s 6.33 and a hallucination score of 7.33 against 2.83, reflecting Gemini's reliance on speculative follow-up framing not supported by the uploaded packet. In Study~3 (kagome-metal mechanistic synthesis), the two systems agreed on the leading hypothesis (fluctuation-mediated pairing near a suppressed CDW/nematic instability), but separated sharply on hallucination (8.50 vs 4.07), because Gemini introduced transport increments and elastoresistance amplitudes that did not appear in the supplied data and presented the strange-metal interpretation as essentially established. In Study~2 (FMAC reproduction), where the task fixes a narrow numerical objective, the two systems performed comparably on task completion (9.00 vs 8.33) and scientific validity (9.00 vs 8.33); {\AgentName} retained a small edge on hallucination (2.17 vs 2.33) by tying its claims more tightly to the computed artefacts.

A representative {\AgentName} execution trajectory on Study~2 is shown in Figure~\ref{fig:real_world}d. The Planner alternates Worker invocations with planning updates: an initial Document-worker extraction of the FMAC theoretical structure is followed by parameter setup and a first solver implementation; intermediate observation flags missing parameter settings and fractional winding indices signalling an incorrect FMAC implementation; the Planner replans, the Worker re-implements under the correct FMAC constraints, and the Critic verifies bulk correspondence and zero-mode multiplicity before the final research note is drafted. This trajectory illustrates that the validity advantage observed in the open-ended studies is accompanied by explicit recovery from model deviation during execution rather than by one-shot answer generation.


Taken together, the user study indicates that on realistic, evidence-bounded physical-science tasks the two systems are closer in nominal task completion than in scientific reliability: {\AgentName} consistently held an advantage on whether final answers remained anchored in the supplied evidence and whether unsupported extrapolation was avoided, with the largest gaps emerging precisely in the studies where the supplied evidence is sparse enough to invite speculative completion (Studies~1 and~3).

\section{Discussion}\label{sec12}

Deep research agents hold genuine promise for accelerating physical science research, yet their capabilities in this domain have remained uncharacterised. Our work addresses this gap with two complementary contributions: {\BenchName}, the first comprehensive benchmark for evaluating deep research capabilities in physical science, and {\AgentName}, a modular multi-agent framework whose architectural specialisation, rather than model scale alone, yields consistent accuracy improvements across diverse scientific tasks. Together, these contributions demonstrate that reliable autonomous scientific reasoning requires purpose-built cognitive architectures grounded in domain knowledge, not merely larger or longer-context foundation models. Crucially, the principal contribution of {\AgentName} is not approaching task saturation on {\BenchName}, but rather shifting the system's dominant failure regime: the coordinated action of planning, memory, and reflection substantially eliminates the coarse-grained agentic instabilities that dominate baseline error profiles, advancing the residual failure frontier to finer-grained scientific grounding and fidelity challenges that demand fundamentally different solutions.

Ablation studies illuminate why architectural specialisation matters. Disabling the Dual-Granularity Memory ($-$3.0\%) or the Hierarchical Physics-Grounded Reflection ($-$3.0\%) each produced the largest single-component degradation, indicating that knowledge grounding and physics-aware validation are the dominant contributors to performance on {\BenchName}. Removing the adaptive planning loop reduced accuracy by 2.5\%, confirming that dynamic replanning remains essential for navigating the long-range dependencies inherent in physical science workflows: without it, early-stage errors in crystal-structure identification or experimental-condition extraction cascade irreversibly through downstream inference. Notably, removing all three components simultaneously reduced accuracy by 6.5\%, less than the sum of individual ablations (8.5\%), revealing partial redundancy in which each mechanism can partially compensate for the others, yet substantially exceeding any single-component removal; the stripped-down pipeline narrows the gap with the strongest generalist baseline from 7.5~p.p.\ to 1.0~p.p., showing that the advantage of {\AgentName} is produced by architectural specialisation rather than by the underlying foundation model. This suggests that grounding large language models in curated domain facts and transferable planning strategies~\cite{boiko2023coscientist, bran2024chemcrow} facilitates reliable performance on complex tasks without requiring specialised parameter tuning, and that the closed cognitive loop derives its strength from the interplay of all three mechanisms rather than from any single component in isolation.

{\BenchName} establishes a rigorous evaluation protocol that may serve as a template for other scientific disciplines~\cite{phan2025hle, mirza2025chembench, laurent2024labbench, tian2024scicode}. By requiring domain experts to originally craft every question from recent publications and adapted textbook problems, the benchmark minimises training-data contamination at the point of creation rather than relying solely on temporal recency. The three-stage cognitive stratification (Information Extraction, Scientific Understanding, and Application and Creation) provides a principled framework for diagnosing which cognitive capabilities limit current systems, an approach readily transferable to chemistry, biology, and materials science benchmarks.

Several limitations warrant consideration. First, the benchmark size of 200 questions, while sufficient to reveal statistically significant performance differences, reflects the high cost of expert curation; community-driven expansion will be essential to broaden domain and difficulty coverage. Relatedly, although the benchmark is balanced across physics and chemistry, coverage within each discipline is deliberately depth-first, concentrated on the most active contemporary subfields (condensed-matter and materials physics; synthetic organic chemistry and catalysis) and sampling adjacent subfields more sparsely, so that aggregate scores should be read as representative of these core areas rather than of uniform disciplinary breadth. Second, evaluating extended scientific answers via large language models may not capture all domain nuances~\cite{krumdick2025no, szymanski2025limitations}, although our blinded validation study demonstrates strong alignment with expert judgment (Spearman $\rho = 0.80$). Third, current vision-language models remain a bottleneck for multimodal understanding~\cite{alampara2025probing,liutowards}, as reflected by a 32\% error rate in this category. Notably, not all residual errors are irreducible: a substantial fraction involves evidence binding and formal-specification fidelity bottlenecks that stronger typed orchestration and validation gates could further reduce, whereas the portion that genuinely reflects foundation-model capability boundaries concentrates in specialised figure interpretation and cross-document synthesis. Fourth, the domain knowledge bases are progressively built from papers retrieved during problem solving and therefore cannot guarantee completeness; extending coverage to additional sub-disciplines and experimental modalities is an important direction for future work.

As autonomous systems become increasingly capable, frameworks like {\AgentName} are designed to assist rather than replace human researchers~\cite{gottweis2025coscientist, lu2024aiscientist, boiko2023coscientist}. We encourage critical human oversight of all content produced by autonomous agents~\cite{resnik2026autonomous, resnik2025ethics}, particularly for sensitive scientific applications where errors carry real-world consequences. Future efforts will focus on expanding {\BenchName} through community contributions, improving multimodal understanding for specialised scientific visualisations, and extending this evaluative framework to additional scientific domains.

\section{Methods}\label{sec11}

\FloatBarrier

\subsection{Benchmark Construction}\label{subsec:benchmark_construction}

To bridge the gap between standardized evaluation and authentic scientific inquiry, {\BenchName} is designed around four core principles: real-world grounding, strict anti-contamination, comprehensive task coverage, and discriminative difficulty. The dataset curation followed a rigorous multi-stage pipeline (Figure~\ref{fig:dataset_overview}a) in which domain experts originally crafted questions drawing on two complementary sources: recent high-impact physical-science publications from the past five years, and problems adapted from authoritative domain textbooks. Because all questions are expert-authored rather than directly copied from existing material, training-data contamination is minimised at the point of creation; the recency of source publications and the reformulation of textbook problems provide additional safeguards. All candidate questions subsequently underwent zero-shot testing against baseline models, with items solvable via parametric memory or general web search excluded from the final set. The final curated benchmark comprises 200 questions, balanced by design between physics and chemistry (100 each), partitioned into six mutually exclusive task categories (Multimodal QA, Long-context QA, Structured Information Extraction, Scientific Reasoning, Experimental Design and Code Generation). To characterise topical coverage without resorting to an ad hoc scheme, each question is additionally assigned a subfield drawn from recognised community classifications: the American Physical Society Physics Subject Headings (PhySH) and the corresponding Physical Review section structure for physics, and the American Chemical Society and IUPAC divisional taxonomy for chemistry.

Because {\BenchName} spans heterogeneous task types, we used a task-specific evaluation protocol rather than a single scoring rule. This protocol combines deterministic rule-based checks with rubric-guided semantic judgment. Structured extraction tasks are scored with a weighted rule-based metric that measures syntactic validity, key match and value match for JSON or CSV outputs. Code-generation tasks are first executed in a sandboxed environment, and the resulting outputs are then evaluated against reference executions using either exact comparison or rubric-guided judgment, depending on the expected answer form. Open-ended responses in Multimodal QA, Long-context QA, Scientific Reasoning and Experimental Design are scored by a rubric-based language-model judge that evaluates each criterion separately and aggregates weighted criterion scores. Short atomic answers are first checked by exact match, and non-matching cases are then adjudicated by a language-model judge against the reference answer. Validation of the composite protocol against independent expert judgment is presented in Figure~\ref{fig:model_analysis}b.

The six categories span the principal cognitive operations of physical-science deep research: \textbf{Multimodal QA} probes perception of scientific figures, \textbf{Long-context QA} probes synthesis across full documents and their supplementary materials, \textbf{Structured Information Extraction} requires schema-conformant parsing into JSON or CSV records, \textbf{Scientific Reasoning} requires multi-step principle-grounded derivation, \textbf{Experimental Design} requires construction of procedurally complete synthesis or characterisation protocols, and \textbf{Code Generation} requires executable computational implementations of physical or chemical systems.

\subsection{{\AgentName} Architecture}\label{subsec:architecture_detailed}

The {\AgentName} architecture, introduced in the Results section and illustrated in Figure~\ref{fig:results_overview}a, comprises a central Planner that orchestrates five specialized worker agents. Below we detail the three core components that form its closed cognitive loop.

To directly operationalize the deep research requirements and ensure execution stability, the architecture is grounded in three tightly coupled components (\textit{Dual-Granularity Memory}, an \textit{Adaptive Planning Loop}, and \textit{Hierarchical Physics-Grounded Reflection}) that together form a closed cognitive loop:

\textbf{Adaptive Planning Loop.} The system operates through a structured execution loop. The Planner dynamically formulates sub-tasks, assigns them to specialized workers for logical deduction, and executes external tool invocations. Crucially, the loop replans in response to intermediate outcomes (including worker failures, unexpected results, and Critic rejections) rather than committing irreversibly to an initial strategy. To maintain a stable working memory state and prevent context degradation during extended reasoning trajectories, a Summarizer module compresses the execution history every three steps, ensuring the Planner retains precise state representations without prompt overflow.

\textbf{Dual-Granularity Memory.} To promote knowledge transferability, the framework maintains memory at two distinct resolutions. At the planning level, an Experience Memory archives successful planning trajectories, dynamically retrieving the top three most similar historical strategies to guide current task decomposition. At the execution level, specialized Domain Knowledge Bases (progressively built from scientific papers retrieved during problem solving) supply the worker agents with reaction mechanisms and experimental protocols.

\textbf{Hierarchical Physics-Grounded Reflection.} The system ensures physical and logical consistency by validating outputs at two levels. At the execution level, domain-specific Verifiers evaluate intermediate worker outputs for fundamental errors, such as mass imbalance or dimensional inconsistency, triggering local refinement loops before escalating failures. At the planning level, a Critic agent evaluates the proposed final answers against a structured error taxonomy, scrutinizing source grounding, chemical logic, and format compliance before terminating the workflow. Verified successful trajectories are fed back into the Experience Memory, closing the cognitive loop.

\subsection{Experimental Setup}\label{subsec:experimental_setup}

To rigorously evaluate the proposed {\AgentName} framework, we established an experimental protocol comparing its performance against a comprehensive suite of state-of-the-art baselines across multiple challenging scientific datasets.

\textbf{Evaluation Datasets.} To assess generalization across diverse scientific domains, we selected three established expert-level benchmarks, specifically isolating their physics and chemistry subsets to align with physical science workflows. First, we utilized the Human's Last Exam (HLE) benchmark~\cite{phan2025hle}, selecting 395 closed-ended questions (230 physics, 165 chemistry) structured to resist superficial retrieval strategies. Second, we incorporated the deep research component of SGI-Bench (SGI-DR)~\cite{sgibench}, evaluating 43 expert-curated questions (32 physics, 11 chemistry) designed to measure iterative reasoning capabilities over complex scientific content. Finally, we assessed open-ended research proficiency using the FrontierScience-Research (FS-Research) track~\cite{frontiersci}, sampling 40 original tasks (20 physics, 20 chemistry). These FS-Research tasks are authored by domain experts and evaluated using a comprehensive 10-point rubric emphasizing scientific judgment.

\textbf{Model Configurations.} We evaluated nine base models spanning proprietary and open-source families (including GPT-5.2~\cite{openai_gpt52}, Gemini 3 Flash/Pro~\cite{google2025gemini3flash, google_gemini3_pro}, Claude Opus 4.5~\cite{anthropic_claude_opus45}, Grok-4.1-Fast~\cite{xai2026grok41fast}, Kimi-K2.5~\cite{moonshot_kimi_k25}, Intern-S1-Pro~\cite{DBLP:journals/corr/abs-2603-25040}, DeepSeek-V3.2-Speciale~\cite{deepseek_v32_speciale} and Qwen3-VL-Plus~\cite{qwen2025qwen3vl}), alongside agentic systems that include Gemini Deep Research (\texttt{deep-research-pro-preview-12-2025})~\cite{google2025geminideepresearch}, OpenAI Deep Research o4-mini~\cite{openai2025deepresearch}, ODR-smolagents~\cite{smolagents}, X-Master~\cite{xmaster}, and Tongyi DeepResearch~\cite{tongyidr}. For {\AgentName}, we uniformly deployed \texttt{gemini-3-flash-preview} as the foundation model across the Planner and all worker agents.

\textbf{Implementation and Hyperparameters.} The multi-agent orchestration of {\AgentName} is implemented with reference to the open-source codebases of Youtu-Agent~\cite{youtu_agent} and OWL~\cite{owl-workforce}. The generation temperature for {\AgentName} was set to 1.0 following official model recommendations for complex reasoning tasks. Specialized tool execution and evaluation scripts were deployed on compute nodes equipped with NVIDIA A800 GPUs. Code generation and computational tasks were executed within isolated Docker sandboxes with strict timeout limits.

\textbf{Evaluation Metrics.} We report overall accuracy (percentage of correctly answered questions) as the primary metric across all benchmarks, scored as pass@1 with a single sampled response per question. For {\BenchName}, correctness is determined by the composite evaluation protocol described above. For the three public benchmarks, we adopt each benchmark's native scoring methodology and run the official evaluation code released with each benchmark: a model-graded protocol (LLM-as-judge) for HLE, exact match for SGI-DR, and the 10-point expert rubric for FS-Research.

\textbf{Validation against expert grading.} A blinded validation set was constructed from the full {\BenchName} and graded by two domain experts (one for physics and one for chemistry, each scoring items within their respective discipline) using a continuous 0--1 rubric. Items for which any automatic metric was undefined were excluded from per-metric correlation analyses; in particular, lexical metrics (BLEU, ROUGE-1/2/L, exact-match) are not defined for the four structured-information-extraction items whose reference outputs are JSON records, yielding $n = 196$ paired items for cross-metric comparison. Agreement between each automatic metric and the expert score was quantified using Spearman's rank correlation $\rho$. Two-sided $p$-values were derived from the exact relation $t = \rho\sqrt{(n{-}2)/(1{-}\rho^{2})}$, and 95\% confidence intervals from Fisher's $z$-transformation, $\rho_{\text{lo,hi}} = \tanh(z \pm 1.96/\sqrt{n{-}3})$ with $z = \tfrac{1}{2}\ln[(1{+}\rho)/(1{-}\rho)]$. The single-LLM-judge baseline (Figure~\ref{fig:model_analysis}b) is implemented with GPT-5.2 acting as a domain-expert grader and returning a binary 0/1 verdict on each (question, reference answer, model response) triple.

\section*{Acknowledgements}
This work is supported by Intern-Discovery.

\clearpage

\begingroup
\sloppy
\printbibliography[heading=bibintoc]
\endgroup

\clearpage

\end{document}